\documentclass[journal,twoside,web]{ieeecolor}
\usepackage{generic}
\usepackage{cite}
\usepackage{amsmath,amssymb,amsfonts}
\usepackage{bm}
\usepackage{booktabs}
\usepackage{algorithmic}
\usepackage{graphicx}
\usepackage{algorithm,algorithmic}
\usepackage{stfloats}
\usepackage{hyperref}
\hypersetup{hidelinks=true}
\usepackage{balance}
\usepackage{textcomp}
\usepackage[table,xcdraw]{xcolor}
\usepackage{graphicx}
\usepackage{xcolor}
\usepackage{multirow}
\usepackage{subfigure} 
\def\BibTeX{{\rm B\kern-.05em{\sc i\kern-.025em b}\kern-.08em
    T\kern-.1667em\lower.7ex\hbox{E}\kern-.125emX}}
\begin{document}
\title{Domain Generalization for Zero-calibration BCIs with Knowledge Distillation-based Phase Invariant Feature Extraction}
\author{Zilin Liang, Zheng Zheng, Weihai Chen, \IEEEmembership{Member, IEEE}, \\Xinzhi Ma, Zhongcai Pei, and Xiantao Sun
\thanks{
This work was supported in part by the Key Research and Development Program of Zhejiang Province under Grant
2021C03050; in part by the Scientific Research Project of Agriculture and Social Development of Hangzhou under Grant 20212013B11; and in part by the National Natural Science Foundation under Grant U1909215, Grant 51975002, and Grant 51975029. (Corresponding author: Weihai Chen; e-mail: whchen@buaa.edu.cn).}
\thanks{
Zilin Liang, Zheng Zheng, Weihai Chen, Xinzhi Ma and Zhongcai Pei are with the School of Automation Science and Electrical Engineering, Beihang University, Beijing 100191, China, 
and also with the Hangzhou Innovation Institute, Beihang University, Hangzhou 310052, China.}
\thanks{
Xiantao Sun is with the School of Electrical Engineering and Automation, Anhui University, Hefei 230601, China.}
}

\maketitle

\begin{abstract}
The distribution shift of electroencephalography (EEG) data causes poor generalization of brain-computer interfaces (BCIs) in unseen domains.
Some methods try to tackle this challenge by collecting a portion of user data for calibration. However, it is time-consuming, mentally fatiguing, and user-unfriendly. 
To achieve zero-calibration BCIs, most studies employ domain generalization (DG) techniques to learn invariant features across different domains in the training set. However, they fail to fully explore invariant features within the same domain, leading to limited performance.
In this paper, we present an novel method to learn domain-invariant features from both inter-domain and intra-domain perspectives.
% The key lies in how to extract intra-domain invariant features within EEG.
% Inspired by the fact that Fourier phase information contains intrinsic semantic and is less susceptible to distribution shift.
For intra-domain invariant features, we propose a knowledge distillation framework to extract EEG phase-invariant features within one domain. 
As for inter-domain invariant features, correlation alignment is used to bridge distribution gaps across multiple domains.
Experimental results on three public datasets validate the effectiveness of our method, showcasing state-of-the-art performance. 
To the best of our knowledge, this is the first domain generalization study that exploit Fourier phase information as an intra-domain invariant feature to facilitate EEG generalization. 
More importantly, the zero-calibration BCI based on inter- and intra-domain invariant features has significant potential to advance the practical applications of BCIs in real world. (The code is available on https://github.com/ZilinL/KnIFE).
\end{abstract}

\begin{IEEEkeywords}
EEG, BCI, Domain Generalization, Transfer Learning, Knowledge Distillation, Phase Information. 
\end{IEEEkeywords}

\section{Introduction}
\label{sec:introduction}
\IEEEPARstart{T}{he} brain-computer interfaces (BCIs) facilitate direct communication between the brain and external devices, circumventing traditional neural pathways \cite{kennedy2000direct}. Electroencephalography (EEG), a safe and portable non-invasive technology, is extensively utilized for recording the brain's neural signals. 
The decoding and analysis of EEG signals enable a wide range of applications, spanning from emotion recognition and external device control to brain fatigue detection, epilepsy monitoring, text output, and beyond. 
These applications hold significant importance in medical, rehabilitation engineering, robotics, and other fields \cite{Zhong2023tactile} \cite{Chowdhury2018Active}.

\begin{figure*}
  \centering
  \includegraphics[scale=1]{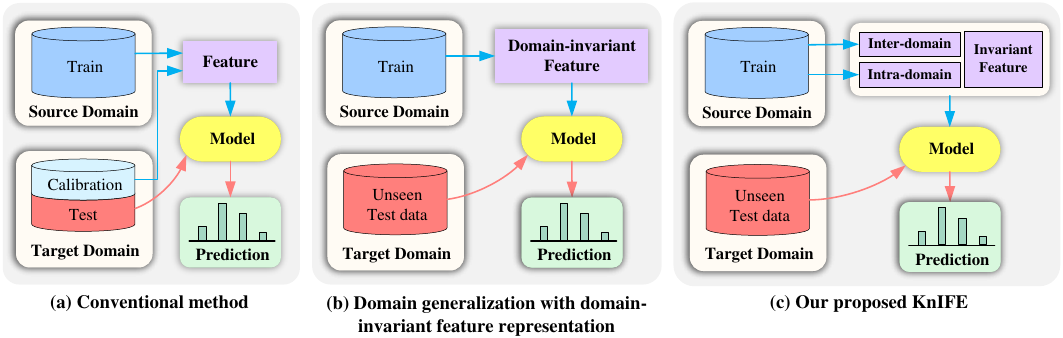}
  \caption{Comparison of conventional method, domain generalization and our proposed method. (a) Conventional methods train on source domain data and calibrate with a portion of target domain data to bridge domain distribution gaps. (b) Domain generalization exclusively relies on source domain data to learn domain-invariant features and generalize to unseen domains. (c) Our approach extends this by considering diverse domain-invariant features, learning them from both intra- and inter-domain viewpoints.}
  \label{Methodcompare}
\end{figure*}

An important challenge in decoding EEG signals is the degradation of system decoding performance caused by distribution shift. This phenomenon arises due to the non-stationary nature of EEG signals, resulting in distribution discrepancies not only among subjects but also within the same subject at different time periods. In methods employing machine learning for EEG decoding, this violates the assumption that data must be independent and identically distributed (i.i.d.). The distribution discrepancies lead to the performance decay of well-trained BCIs when applied to new subject. Some approaches attempt to collect additional data from new subjects for model calibration \cite{he2019transfer}, as shown in Fig. \ref{Methodcompare}(a). Nevertheless, EEG data collection is time-consuming, rendering this additional time requirement unfriendly and inconvenient for users. Although domain adaptation is employed to handle the distribution shift, it still require a certain amount of unlabeled target domain data for training \cite{zhu2021deep, liu2023multi, jiao2019sparse, azab2019weight, liang2024manifold}. For practical BCI usage, obtaining EEG data from new BCI users is usually not feasible.

Recently, domain generalization (DG) has been proposed to address this more challenging and practical problem. DG aims to achieve robust generalization performance in unseen domains by leveraging knowledge acquired only from the source domain \cite{zhou2023domain}, as shown in Fig. \ref{Methodcompare}(b). Domain-invariant feature learning, renowned for its effectiveness in DG, focuses on representations of domain-invariant features within a specialized feature space across multiple source domains. The domain-invariant feature is consistent in the source and unseen domains, so it can achieve excellent generalization from source domains to unseen domains.
Based on this advantage, the DG is used in BCIs to achieve cross-subject/session generalization \cite{han2024noise} \cite{kwak2023subject}.
However, these methods only improve the generalization from the perspective of inter-domain invariant features.
Some recent studies pointed that simple feature alignment or inter-domain invariant features may be insufficient for generalization \cite{bui2021exploiting} \cite{zhao2019learning}.
Zhao et al. \cite{zhao2019learning} proposed a disentanglement framework to learn invariant and specific features. Lu et al. \cite{lu2022domain} learned domain-invariant representations from internal and mutual sides.
These researches suggest that exploring the diversity of invariant features enhances generalization, although mainly focused in the computer vision field.
Inspired by these researches, our study focused on extracting invariant features from both inter- and intra-domain perspectives, as shown in Fig. \ref{Methodcompare}(c).
More specifically, the intra-domain invariant features capture intrinsic knowledge from data in its own domain not affected by data in other domains. The inter-domain invariant features aim to identify transferable knowledge across multiple domains.

The key challenge is how to extract the intra-domain invariant features of EEG data.
More recently, researches suggest that the phase component in the Fourier spectrum preserves high-level semantic information from the original signal and is less susceptible to domain shifts, which may be a kind of intra-domain invariant features \cite{yang2020phase,oppenheim1981importance,lin2023deep}. 
It is not easy to extract phase-invariant features on EEG, although it has recently been validated on images.
The main technical challenges are: (1) It is more difficult to extract invariant features from EEG because EEG signals are dynamically changing and highly susceptible to interference. The low signal-to-noise ratio of EEG makes the phase information extracted by traditional Fourier transform contain large errors.
(2) Unlike images that are composed of static pixels, EEG signals change non-steadily and dynamically along the time axis, and the phase information is submerged in various amplitudes and noise. Moreover, the EEG dataset has a smaller number of samples than the image dataset, which makes it more difficult to extract phase information in a deep neural network. 
It remains an open problem to develop an effective approach to extract phase-invariant features from EEG data. 

To address the above issues, we propose \underline{Kn}owledge Distillation-based Phase \underline{I}nvariant \underline{F}eature \underline{E}xtraction framework (KnIFE), to “dissect” the intrinsic invariant features hidden in the EEG data. 
It is a plug-and-play zero-calibration method, which utilizes pre-collected EEG data for model training and can be directly applied to new users.
For intra-domain invariant features, the KnIFE uses knowledge distillation to extract Fourier phase-invariant features. For inter-domain invariant features, correlation alignment is used to bridge distribution gaps between any two domains. 
To the best of our knowledge, this is the first domain generalization study that exploit phase-invariant feature to facilitate EEG generalization.
The main contributions are:

1) \emph{Diverse domain-invariant features}: we propose a domain generalization method that simultaneously considers inter- and intra-domain invariant features to enhance EEG generalization capability.

2) \emph{Fourier phase-invariant feature extraction}: we explore Fourier phase-invariant features as intra-domain invariant features and propose a knowledge distillation framework to extract them from EEG signals.

3) \emph{Spectrum transfer}: we introduce a Fourier-based spectrum transfer method for data augmentation, which highlight phase information in EEG signal processing.

The remainder of this paper is organized as follow: Section II introduces related work on domain generalization, knowledge distillation, and Fourier phase features. Section III details the proposed KnIFE. The experimental results are presented in section IV. In section V, the proposed method is discussed in detail and the future research direction is pointed out. Finally, section VI concludes the paper. 

\section{Related Work}

\subsection{Domain Generalization}
The goal of domain generalization is to train a model on multiple distinct yet related source domains, enabling effective generalization to entirely new and unseen target domains.
This presents a more challenging yet applicable scenario for real-world applications.
Domain-invariant representation learning stands out as the predominant approach in domain generalization. 
Ben et al. \cite{ben2006analysis} theoretically demonstrates that if feature representations remain invariant across different domains, they are general and transferable between domains.
Existing studies primarily learn domain-invariant features through simple alignment or by reducing distribution discrepancies between any two domains \cite{borgwardt2006integrating} \cite{zhu2021deep} \cite{sun2016return} \cite{liu2023multi}. These studies employ technologies such as maximum mean discrepancy (MMD), local maximum mean discrepancy (LMMD) and Coral. 
However, these studies only learn inter-domain invariant features, which might not suffice for effective generalization \cite{zhao2019learning}.
Thus, exploring a more diverse set of invariant features is crucial for advancing domain generalization techniques.

\subsection{Knowledge Distillation}
Initially proposed by Hinton et al. \cite{hinton2015distilling} for model compression, knowledge distillation involves transferring knowledge from a powerful and complex model (teacher) to a lighter and more deployable model (student) designed for a specific task. The teacher network is typically a deep model with an extensive number of parameters trained on massive datasets. In contrast, the student network is a lightweight model guided by teacher network’s soft labels. Knowledge distillation empowers the student network to achieve performance comparable to that of the teacher network on specific tasks. 
The 'teacher-student' framework has demonstrated effectiveness in numerous application scenarios \cite{wangselfguided} \cite{rahimpourcross}.
Some studies have applied it to domain generalization for learning domain-invariant features, particularly in extracting phase information from images\cite{xu2021fourier}\cite{lu2022domain}.
In this study, knowledge distillation is explored to delve deeply into the hidden phase-invariant features embedded within EEG data.

\begin{figure*}
  \centering
  \includegraphics[scale=1]{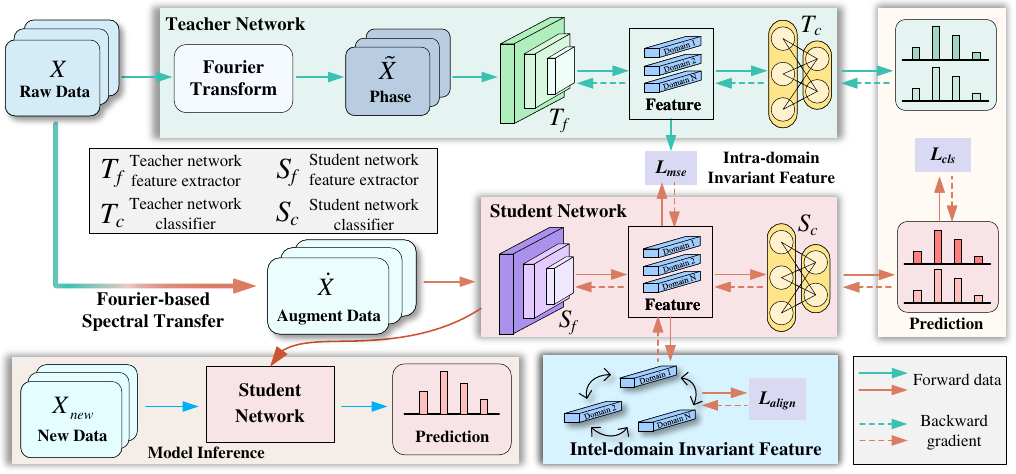}
  \caption{The detailed framework of KnIFE. Firstly, the teacher network is responsible for learning the phase information of the raw data and transferring this knowledge to the student network. This process enables the student network to effectively extract intra-domain invariant features. To facilitate this, Fourier-based spectral transfer techniques are employed to emphasize phase information. Additionally, the student network learns inter-domain invariant features by aligning the distribution between any pair of domains. During the model inference stage, the new subjects' data can be directly fed into the pre-trained student network model for prediction.}
  \label{frameworkFoSTer}
\end{figure*}

\subsection{Fourier Phase Features}
Fourier phase analysis has emerged as a promising technique for domain generalization, as demonstrated in recent studies \cite{xu2021fourier} \cite{yang2020fda}. Xu et al. \cite{xu2021fourier} observed that images' phase information carries high-level semantic content that remains robust against domain shifts, leading to effective generalization to unseen domains. 
Lu et al. \cite{lu2022domain} further confirmed that Fourier phase features serve as internally invariant features, enhancing domain generalization performance. 

However, applying Fourier phase analysis to EEG data presents unique challenges.
EEG signals, being time series data, exhibit chaotic variations along the time axis, complicating the extraction of phase features.
Factors such as susceptibility to interference, low signal-to-noise ratio, limited sample size, and submergence of phase information under interference exacerbate this challenge. 
To address this, we leverage spectral transfer to suppress amplitude variations and accentuate phase information.
% This approach aims to mitigate the impact of interference and noise, facilitating the extraction of robust domain-invariant features for improved generalization across diverse EEG datasets.

\section{Methodology}  
In this section, we elaborate on the proposed method, as shown in Fig. \ref{frameworkFoSTer}.
%\footnote{The code is available on https://github.com/ZilinL/KnIFE}

\subsection{Problem Definition}
The training set consists of multiple labeled source domains, denoted as $\mathcal{D}_s = \left\{ \mathcal{D}_1, \ldots, \mathcal{D}_N \right\}$, each containing $N_i$ labeled samples $ \left\{ (x_i,y_i) \right \}_{i=1}^{N_i}$ in the $i$th domain $\mathcal{D}_i$. Here, $x_i \in \mathbb{R}^{n\times m }$ represents multi-channel EEG data with $n$ channels and $m$ sampling points, and $y_i$ represents the corresponding labels. Our goal is to train a model, denoted as $y=f(x)$, using pre-existing source domain data to learn domain-invariant feature expressions. This facilitates robust and generalized classification on unseen target domains $\mathcal{D}_t$.

\begin{figure}
  \centering
  \includegraphics[scale=1]{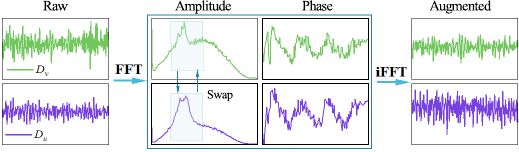}
  \caption{Diagram of spectral transfer. $\mathcal{D}_v$ and $\mathcal{D}_u$ represent the EEG data of two subjects. FFT and iFFT represent the fast Fourier transform and the inverse transform, respectively.}
  \label{Spectral_Transfer}
\end{figure}

\subsection{Fourier-based Spectral Transfer}
Spectral transfer maps the source domain data to the style of the target domain without altering semantic content \cite{xu2021fourier}. In EEG data, it enhances phase information by disturbing amplitudes, as phase information inherently contains domain-invariant features.

Consider a single-channel $N$-point discrete sequence of EEG data, denoted as $x$. The Fourier transform $\mathcal{F}(x)$ is expressed as:
\begin{equation}
    \mathcal{F}(x)(k)= \sum_n x(n)e^{-i2\pi k n/N}
    \label{FourierTransfer}
\end{equation}
where $k$ is the index.
The amplitude and phase components are respectively expressed as:
\begin{equation}
    \mathcal{A}(x)(k)= \left[R^2(x)(k) + I^2(x)(k)\right] ^ {1/2}
\end{equation}

\begin{equation}
    \mathcal{P}(x)(k)=\arctan  \left[  \frac{I(x)(k)}{R(x)(k)} \right]
\end{equation}
where $R(x)$ and $I(x)$ denote the real and imaginary parts of $\mathcal{F}(x)$, respectively.

Previous studies suggest that the phase information is less susceptible to domain drift. Therefore, we focus on extracting domain-invariant features from the phase by fixing it and swapping the amplitude component, as illustrated in Fig. \ref{Spectral_Transfer}. Additionally, we introduce a mask $M_{\alpha}(n)$ to specify the range of the amplitude swap:
\begin{equation}
M_{\alpha}(n)=\left\{
\begin{array}{l}
    1, n \in [- \alpha N, \alpha N] \\
    0, others \\
    \alpha \in (0,0.5)
\end{array}
\right.
\end{equation}

Given randomly sampled EEG data $\boldsymbol{x}^u \sim \mathcal{D}_u$ and $\boldsymbol{x}^v \sim \mathcal{D}_v$, where $\mathcal{D}_u$ and $\mathcal{D}_v$ represent two different domains, the data augmented by spectral transfer can be represented as:
\begin{equation}
\begin{split}
    \dot{\boldsymbol{x}}^{u\rightarrow v} &= \mathcal{F}^{-1}\left(M_{\alpha}\circ \mathcal{A}(\boldsymbol{x}^v)+(1-M_{\alpha})\circ \mathcal{A}(\boldsymbol{x}^u), \mathcal{P}(\boldsymbol{x}^u)\right) \\
    \dot{\boldsymbol{x}}^{v\rightarrow u} &= \mathcal{F}^{-1}\left(M_{\alpha}\circ \mathcal{A}(\boldsymbol{x}^u)+(1-M_{\alpha})\circ \mathcal{A}(\boldsymbol{x}^v), \mathcal{P}(\boldsymbol{x}^v)\right)
\end{split}
\end{equation}
where $\mathcal{F}^{-1}(\mathcal{A}(x),\mathcal{P}(x))$ denotes the inverse Fourier transform.

\subsection{Knowledge Distillation of Phase Information}
We establish a knowledge distillation framework to extract the intra-domain invariant feature from Fourier phase information.
As illustrated in Fig. \ref{frameworkFoSTer}, the raw data is transformed into phase components by the Fourier transform (Eq. (\ref{FourierTransfer})) within the teacher network. Subsequently, phase components undergo feature extraction and classification within the network. Denoting $\widetilde{\boldsymbol{x}}$ as the phase of $x$, the training data for the teacher network becomes $(\widetilde{\boldsymbol{x}},y)$, and its training process is formulated as:

\begin{equation}
    \min_{\theta_T^f,\theta_T^c}\mathbb{E}_{\left(\widetilde{\boldsymbol{x}},y\right)\backsim P^{tr}} \mathcal{L}_{cls}^T(T_c(T_f(\widetilde{\boldsymbol{x}};\theta_T^f);\theta_T^c),y),
\end{equation}
where $\theta_T^f$ and $\theta_T^c$ represent the training parameters of the feature extractor $T_f$ and classifier $T_c$ within the teacher network, respectively. $\mathbb{E}$ and $P^{tr}$ denote the expectation and distribution of training data, respectively. $\mathcal{L}_{cls}$ is the cross-entropy loss function.

After obtaining a pre-trained teacher network model, we utilize it to guide the student network to extract phase-invariant features.
The augmented data is denoted as $\dot{x}$, and the training data for the student network is $(\dot{\boldsymbol{x}},y)$.
The teacher network guides the student network through knowledge distillation, which can be expressed as:
\begin{equation}
\begin{split}
    \min_{\theta_S^f,\theta_S^c} & \mathbb{E}_{\left(\widetilde{\boldsymbol{x}},y\right)\backsim P^{tr}}\mathcal{L}_{cls}^S(S_c(S_f(\widetilde{\boldsymbol{x}};\theta_S^f);\theta_S^c),y) \\ + \gamma_1 & \mathbb{E}_{\left(\dot{\boldsymbol{x}},y\right)\backsim P^{tr}} \mathcal{L}_{mse}(S_f(\dot{\boldsymbol{x}};\theta_S^f),T_f(\dot{\boldsymbol{x}})),
\end{split}
\end{equation}
where $\theta_S^f$ and $\theta_S^c$ represent the training parameters of the feature extractor $S_f$ and classifier $S_c$ within the student network, respectively. $\gamma_1$ is a hyperparameter.
$\mathcal{L}_{mse}$ denotes the mean square error (MSE) loss function, aimed at preserving the similarity between the features extracted by the teacher and student networks.

\subsection{Cross-Domain Invariant Feature Extraction}
In this section, we consider inter-domain invariant feature extraction by exploring Correlation alignment across multiple domains.
Inter-domain invariant features are attained by minimizing the statistical distribution discrepancy between any two domains.
Let $C_v$ denote the covariance matrix of features in domain $D_v$, expressed as:
\begin{equation}
  C_v=\frac{1}{N_v} (V_v^T V_v - \frac{1}{N_v} (1^T V_v)^T (1^T V_v)). \\
\end{equation}
Here, $V_v=S_f(\dot{x}_v)$ represents the features extracted by the student network in the $v$th domain.
For any two distinct domains, we perform second-order statistical correlation alignment, and the loss function is formulated as:
\begin{equation}
  \mathcal{L}_{align} = \frac{2}{N(N-1)} \sum_{v\neq u}^N \frac{1}{4d^2} \left\lVert C_v - C_u \right\rVert ^2_F.
\end{equation}
Here, $\left\lVert \cdot \right\rVert ^2_F $ denotes the squared Frobenius norm, and $d$ indicates the dimensionality of the feature vectors.

\subsection{Overview of loss}
By combining the aforementioned loss functions, the comprehensive loss function for training the student network is expressed as:
\begin{equation}
  \mathcal{L}_S =  \mathcal{L}_{cls}^S + \gamma_1 \mathcal{L}_{mse} + \gamma_2 \mathcal{L}_{align}.
\label{all_loss}
\end{equation}
Here, $\gamma_1$ and $\gamma_2$ serve as tradeoff hyperparameters.
The loss function comprises three terms: 

$\mathcal{L}_{cls}^S$ corresponds to the classification loss, aiming to ensure the accuracy of the student network's classification. 

$ \mathcal{L}_{mse}$ is the MES loss, empowering the student network with the capability to learn phase-invariant features during the knowledge distillation process. 

$\mathcal{L}_{align}$ quantifies the similarity between domains by aligning feature distributions, contributing to the extraction of inter-domain invariant features.

$ \mathcal{L}_{mse}$ and $\mathcal{L}_{align}$ extract domain-invariant features from intra- and inter-domain perspectives, respectively, enabling the model to possess strong generalization abilities and perform well on unseen domain data.

In the inference stage, for the newly arrived unseen domain data $\boldsymbol{x}_{new}$, we predict its label by:
\begin{equation}
  y =  S_c(S_f(\boldsymbol{x}_{new})).
\label{all_loss}
\end{equation}

\section{Experiments and Results}

The effectiveness of our method is validated through experiments and compared with several advanced algorithms.

% Table generated by Excel2LaTeX from sheet 'Sheet2'
\begin{table}[ht]
    \centering
    \caption{Summary of datasets}
    \setlength{\tabcolsep}{3pt}
    \begin{tabular}{cccccc}
    \toprule
    Datasets & \#Subjects & \#Channels & \#Classes & \#Sessions & \#Trials \\
    \midrule
    OpenBMI & 54    & 20    & 2     & 2     & 200*54*2 \\
    BCICIV-2a & 9     & 22    & 4     & 2     & 288*9*2 \\
    BCICIV-2b & 9     & 3     & 2     & 5     & (120$\backsim$160)*9*5 \\
    \bottomrule
    \end{tabular}%
    \label{datasets}%
\end{table}%

\subsection{Dataset}
\subsubsection{OpenBMI Dataset}
This dataset\footnote{http://gigadb.org/dataset/view/id/100542/File\_page} is a large-scale EEG dataset containing data from 54 participants, each involved two distinct motor imagery tasks (left hand and right hand). Each task was repeated 100 times. 
The data consists of 20 channels with a sampling rate of 250 Hz. 
The experiment comprises two sessions conducted on different dates. We utilize the data from the first session for training and the data from the second session for testing.

\subsubsection{BCICIV-2a Dataset}
This dataset\footnote{https://www.bbci.de/competition/iv/\#dataset2a} is a four-class motor imagery dataset derived from dataset 2a of the BCI Competition IV. It includes EEG recordings from 22 channels sampled at 250 Hz and is collected from 9 subjects. Each subject performed four motor imagery tasks: left hand, right hand, tongue, and feet, with 72 trials for each task. The dataset comprises two sessions, one for training and the other for testing.

\subsubsection{BCICIV-2b Dataset}
This dataset\footnote{https://www.bbci.de/competition/iv/\#dataset2b} contains data from 9 subjects recorded from three channels (C3, CZ and C4) with a sampling rate of 250 Hz. Each subject performed two-category motor imagery tasks involving the left and right hands. There are 5 sessions per subject collected on different days, with each session containing 120 to 160 samples. The first three sessions are allocated for training, while the last two sessions are designated for testing.

% Table generated by Excel2LaTeX from sheet 'BCICIV2a'
\begin{table*}[ht]
  \centering
  \caption{Classification accuracy (\%) and standard deviation (STD) on dataset BCICIV-2a.}
    \begin{tabular}{c|ccccccccccc}
    \toprule
    Methods & A1    & A2    & A3    & A4    & A5    & A6    & A7    & A8    & A9    & Avg   & STD \\
    \midrule
    ERM   & 76.74  & 47.92  & 85.07  & 62.50  & 59.03  & \underline{59.38}  & \underline{80.21}  & 75.69  & 65.63  & 68.02  & 11.40  \\
    DANN  & 63.19  & 44.10  & 83.33  & 54.86  & 57.64  & 53.47  & 72.57  & 70.83  & 56.25  & 61.80  & 11.26  \\
    RSC   & \underline{80.90}  & \textbf{53.47} & 85.76  & \textbf{62.85} & 55.21  & 58.68  & 77.43  & \underline{79.17}  & \underline{70.14}  & \underline{69.29}  & 11.42  \\
    Mixup & 74.31  & 47.22  & 86.46  & 61.11  & \textbf{63.89} & 57.99  & 77.78  & 75.00  & 61.11  & 67.21  & 11.37  \\
    MMD   & 62.15  & 33.68  & 75.69  & 39.93  & 34.38  & 57.64  & 51.39  & 60.76  & 60.76  & 52.93  & 13.47  \\
    CORAL & 69.44  & 41.67  & 76.04  & 46.88  & 34.72  & 57.99  & 57.99  & 67.71  & 65.97  & 57.60  & 13.10  \\
    VREx  & 79.17  & \underline{50.69}  & \underline{87.50}  & 59.03  & 63.89  & 56.94  & 77.78  & 74.65  & 65.97  & 68.40  & 11.39  \\
    GroupDRO & 62.85  & 30.56  & 76.39  & 41.67  & 31.60  & 56.25  & 50.00  & 65.28  & 56.94  & 52.39  & 14.59  \\
    MLDG  & 40.63  & 37.85  & 49.31  & 49.65  & 59.38  & 32.29  & 50.35  & 50.00  & 30.56  & 44.45  & 9.06  \\
    MIDRL & 73.61  & 37.85  & 59.72  & 48.26  & 33.33  & 32.64  & 44.44  & 73.26  & 61.81  & 51.66  & 15.16  \\
    SMA   & 73.26  & 36.81  & 52.78  & 41.67  & 33.66  & 31.60  & 49.31  & 69.10  & 61.46  & 49.96  & 14.51 \\ 
    FDCL  & 76.04  & 38.19  & 68.06  & 53.13  & 35.42  & 34.03  & 44.79  & 77.78  & 62.85  & 54.48  & 16.35  \\
    KnIFE (ours) & \textbf{82.29} & 50.35  & \textbf{89.58} & \underline{62.50}  & \underline{63.19}  & \textbf{61.81} & \textbf{81.60} & \textbf{80.90} & \textbf{70.14} & \textbf{71.37} & 12.14  \\
    \bottomrule
    \end{tabular}%
  \label{acc_datasetBCIC}%
\end{table*}%

\begin{table*}[ht]
  \centering
  \caption{Classification accuracy (\%) and standard deviation (STD) on dataset BCICIV-2b.}
    \begin{tabular}{c|ccccccccccc}
    \toprule
    Methods & B1    & B2    & B3    & B4    & B5    & B6    & B7    & B8    & B9    & Avg   & STD \\
    \midrule
    ERM   & 70.00  & 64.29  & 68.44  & \underline{95.31}  & 94.69  & 83.44  & 85.00  & 90.00  & 82.50  & 81.52  & 10.82  \\
    DANN  & 69.69  & 61.07  & 65.63  & 92.81  & 85.63  & 79.38  & 78.44  & 89.38  & 82.81  & 78.31  & 10.22  \\
    RSC   & 70.00  & 60.71  & \textbf{69.38} & \textbf{95.94} & 88.75  & 80.00  & 80.94  & 88.44  & 83.75  & 79.77  & 10.55  \\
    Mixup & 72.81  & 65.71  & \underline{69.38}  & 95.31  & 94.38  & 83.13  & \textbf{86.56} & 89.06  & 82.19  & \underline{82.06}  & 10.06  \\
    MMD   & 73.13  & 64.29  & 67.50  & 89.38  & 93.75  & 77.81  & 80.31  & \underline{91.56}  & 79.69  & 79.71  & 9.81  \\
    CORAL & 74.38  & 65.71  & 66.56  & 91.25  & 93.75  & 82.81  & 81.25  & 87.81  & 79.38  & 80.32  & 9.44  \\
    VREx  & 71.56  & 63.21  & 67.50  & 94.06  & 94.38  & 82.81  & 85.94  & 90.63  & 81.25  & 81.26  & 10.84  \\
    GroupDRO & 73.44  & \underline{66.79}  & 68.44  & 92.19  & 94.69  & 78.44  & 80.31  & 89.06  & 81.88  & 80.58  & 9.43  \\
    MLDG  & 73.44  & 63.93  & 68.75  & 89.38  & \underline{95.31}  & 78.75  & 81.56  & 89.06  & 79.06  & 79.92  & 9.65  \\
    MIDRL & 70.31  & 58.93  & 55.00  & 91.56  & 69.38  & 80.31  & 73.75  & 82.50  & \underline{88.44}  & 74.46  & 11.79  \\
    SMA   & 71.25  & 58.93  & 52.19  & 90.31  & 67.19  & 82.19  & 73.75  & 82.50  & 89.06  & 74.15  & 12.42 \\ 
    FDCL  & \textbf{75.63} & 59.64  & 56.25  & 93.13  & 78.13  & \underline{84.06}  & 77.19  & 82.19  & \textbf{90.00} & 77.36  & 11.74  \\
    KnIFE (ours) & \underline{74.69}  & \textbf{66.79} & 65.31  & 93.75  & \textbf{95.94} & \textbf{85.00} & \underline{85.94}  & \textbf{93.75} & 81.56  & \textbf{82.53} & 10.83  \\
    \bottomrule
    \end{tabular}%
  \label{acc_datasetBCICIV2b}%
\end{table*}%

\subsection{Baselines}
The proposed KnIFE is compared with a variety of baselines.

\textbf{Classical neural networks}

EEGNet \cite{lawhern2018eegnet}: A compact convolutional neural network specifically designed for EEG data analysis. 

ShallowNet: \cite{sakhavi2018learning}: A neural Network with two convolutional layers.

DeepNet \cite{schirrmeister2017deep}: A deep CNN architecture comprising five convolutional layers.

% DANN \cite{ganin2016domain}: A network employing adversarial generative mechanisms to reduce domain discrepancies. 

% Coral \cite{sun2016deep}: A domain method that aligns the second-order statistics of the source and target distributions with a linear transformation.

\textbf{Domain generalization}

Deep Domain Generalization Toolkit (DeepDG): DeepDG is a testbed for domain generalization, offering a fair training and validation protocol for various DG methods.
The toolkit integrates several state-of-the-art DG methods, including: ANDMask \cite{parascandolo2021learning}, Verx \cite{krueger2021out}, GroupDRO \cite{sagawa2019distributionally}, RSC \cite{huang2020self}, DANN \cite{ganin2016domain}, Coral \cite{sun2016deep}, Mixup  \cite{zhang2018mixup} and MLDG  \cite{li2018learning}.

DG methods dedicated to EEG decoding in recent years:
SMA(2022) \cite{arpit2022ensemble} , MIDRL(2023) \cite{jeon2023mutual}, FDCL(2023) \cite{liang2023domain}, DRLF(2024) \cite{han2024noise}.

\subsection{Implementation Details}
\begin{figure}
  \centering
  \includegraphics[scale=1]{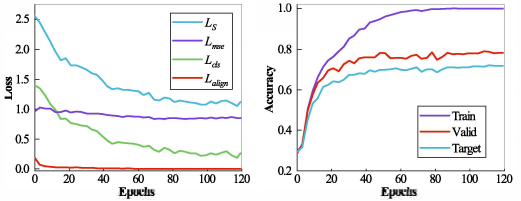}
  \caption{Loss and accuracy vary with epoch during training. All losses gradually decrease and converge during the training process.}
  \label{loss}
\end{figure}

To ensure the applicability of our approach in real-world scenarios, we adhere to strict separation between training and test data, with the latter not involved in the training process. EEG data are band-pass filtered at 4-40 Hz. Within the training set,80\% of the data is allocated for training and the remaining 20\% for validation. The model with the highest performance on the validation set is retained for testing on the test set, yielding the final classification results. The batch size is set to 32 and the model is trained for 200 epochs on OpenBMI dataset and 120 epochs on the other two datasets. Stochastic gradient descent (SGD) is employed as the optimizer with an initial learning rate of 0.005. The learning rate decreases by 0.1 at 70\% and 90\% of the maximum epoch number, respectively. Fig. \ref{loss} shows the convergence of the loss and accuracy during the training on the BCICIV2a dataset.
% The values of both $\lambda_1$ and $\lambda_2$ are set to 1.
The experiments were implemented on a TITAN RTX GPU with 24GB of memory.

\subsection{Quantitative Evaluation}
Table \ref{acc_datasetBCIC}, \ref{acc_datasetBCICIV2b} and \ref{acc_OpenBMI} present the experimental results of our proposed method and the comparison methods on datasets BCICIV2a, BCICIV2b and OpenBMI, respectively. 
The highest accuracy is highlighted in \textbf{bold} and the second-best result is \underline{underlined}. 
Table \ref{acc_datasetBCIC} reveals that our KnIFE method achieved the highest accuracy in 6 out of 9 subjects on dataset BCICIV2a. 
Additionally, KnIFE achieved the second-highest classification accuracy on subjects A4 and A5. Furthermore, the average accuracy of KnIFE surpassed all other methods, outperforming the second-ranking RSC method by 2.08\%. 
In Table \ref{acc_datasetBCICIV2b}, KnIFE achieved the highest accuracy in 4 out of 9 subjects and the second-highest accuracy on subjects B1 and B7 on dataset BCICIV2b. KnIFE's average accuracy was also the highest among all methods, 0.47\% higher than the second-ranked Mixup.

Regarding Table \ref{acc_OpenBMI}, we examined the impact of different backbones in addition to comparing different methods. 
On average, KnIFE exhibited the highest accuracy among the different backnbones, surpassing RSC, which had the second-highest accuracy, by 0.7\%.
This indicates the universality of our method across various backbones. 
Notably, the best performance was achieved when using EEGNet as the backbone, with an accuracy of 83\%.  
However, when ShallowNet is selected as the backbone, KnIFE's performance is inferior to that of EEGNet and DeepNet as the backbone, ranking second among all comparison methods, with a 0.65\% lower accuracy than RSC.

Table \ref{Ttest} displays the p-values of paired t-tests between KnIFE and the comparison algorithms. Values less than 0.05 are bolded. We rounded the results to four decimal places. Notably, KnIFE significantly outperformed other comparison methods on the BCICIV2a dataset, with all p-values being less than 0.05.
In both OpenBMI and the BCICIV2b dataset, only two methods exhibited significance levels higher than 0.05 when compared to KnIFE.

In conclusion, our proposed KnIFE achieves state-of-the-art performance. Its accuracy is significantly higher than other methods. Compared to recent EEG generalization method DRLF, KnIFE exhibits a significant advantage, with an accuracy improvement of 4.11\%. This improvement is attributed to the introduction of phase-invariant feature, which enhances KnIFE's generalization capability from intra-domain invariant feature perspective, as discussed in detail in the section \ref{dis}.

% Table generated by Excel2LaTeX from sheet 'OpenBMI'
\begin{table}[ht]
  \centering
  \caption{Classification accuracy (\%) and standard deviation (STD) of different backbones on dataset OpenBMI}
  \setlength{\tabcolsep}{4pt}
    \begin{tabular}{c|cc|cc|cc|c}
    \toprule
    Backbones & \multicolumn{2}{c|}{EEGNet} & \multicolumn{2}{c|}{DeepNet} & \multicolumn{2}{c|}{ShallowNet} & \multirow{2}[4]{*}{Avg} \\
\cmidrule{1-7}    Methods & Acc   & STD   & Acc   & STD   & Acc   & STD   &  \\
    \midrule
    ERM   & 80.51  & 11.99  & 80.46  & 12.53  & 78.87  & 12.03  & 79.95  \\
    DANN  & 79.00  & 11.24  & 79.47  & 12.15  & 77.19  & 11.92  & 78.55  \\
    RSC   & \underline{82.76}  & 11.20  & 79.07  & 12.93  & \textbf{80.59}  & 11.98  & \underline{80.81}  \\
    Mixup & 82.38  & 11.11  & 80.79  & 12.30  & 79.25  & 11.31  & 80.81  \\
    MMD   & 79.19  & 12.26  & \underline{81.38}  & 11.78  & 78.13  & 12.22  & 79.57  \\
    CORAL & 80.48  & 11.25  & 80.90  & 11.89  & 78.97  & 11.74  & 80.12  \\
    VREx  & 81.04  & 11.58  & 80.56  & 12.21  & 79.03  & 12.11  & 80.21  \\
    GroupDRO & 79.43  & 10.52  & 78.69  & 12.98  & 79.66  & 11.50  & 79.26  \\
    DRLF  & 77.40  & 9.70 & 77.40  & 9.70 & 77.40  & 9.70 & 77.40 \\
    MLDG  & 80.19  & 11.86  & 80.96  & 11.63  & 76.60  & 12.22  & 79.25  \\
    ANDMask & 80.12  & 11.35  & 77.07  & 13.57  & 78.88  & 12.06  & 78.70  \\
    KnIFE (Ours) & \textbf{83.00} & 11.12  & \textbf{81.58}  & 11.89  & \underline{79.94}  & 11.55  & \textbf{81.51} \\
    \bottomrule
    \end{tabular}%
  \label{acc_OpenBMI}%
\end{table}%

% Table generated by Excel2LaTeX from sheet 'OpenBMI'
\begin{table}[ht]
  \centering
  \caption{The p-values in paired t-test ($\alpha=0.05$)}
    \begin{tabular}{cccc}
    \toprule
    \multirow{2}[4]{*}{Methods} & \multicolumn{3}{c}{p-valve} \\
\cmidrule{2-4}          & \multicolumn{1}{l}{BCICIV-2a} & \multicolumn{1}{l}{OpenBMI} & \multicolumn{1}{l}{BCICIV-2b} \\
    \midrule
    ERM   & \textbf{0.0003} & \textbf{0.0000} & 0.1332  \\
    DANN  & \textbf{0.0001} & \textbf{0.0000} & \textbf{0.0051} \\
    RSC   & \textbf{0.0427} & 0.2166  & \textbf{0.0445} \\
    Mixup & \textbf{0.0018} & 0.0508  & 0.2962  \\
    MMD   & \textbf{0.0001} & \textbf{0.0000} & \textbf{0.0069} \\
    CORAL & \textbf{0.0005} & \textbf{0.0000} & \textbf{0.0076} \\
    VREx  & \textbf{0.0023} & \textbf{0.0000} & \textbf{0.0430} \\
    GroupDRO & \textbf{0.0001} & \textbf{0.0000} & \textbf{0.0494} \\
    MLDG  & \textbf{0.0002} & \textbf{0.0001} & \textbf{0.0131} \\
    \bottomrule
    \end{tabular}%
  \label{Ttest}%
\end{table}%

\begin{figure}[ht]
  \centering
  \includegraphics[scale=1]{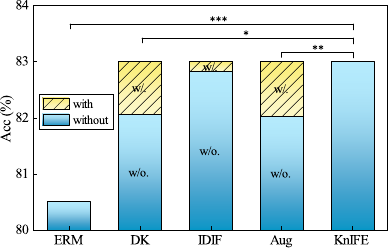}
  \caption{Ablation study of different components on KnIFE.}
  \label{ablation}
\end{figure}

\subsection{Ablation Study}
The ablation study was conducted to assess the individual contributions of each component in KnIFE. We divided KnIFE into three components: knowledge distillation (DK) for intra-domain invariant feature extraction, data augmentation (Aug), and inter-domain invariant feature extraction (IDIF). Experiments were conducted on the OpenBMI dataset by removing one of these components at a time to form new models, and the results are illustrated in Fig. \ref{ablation}. While the new models, each missing one of the three components, outperform the baseline model ERM, they exhibited inferior performance compared to the complete KnIFE model. Table \ref{loss_ablation} further analyzes the influence of each loss component in Eq. (\ref{all_loss}). The model's accuracy surpasses that of the baseline when solely $\mathcal{L}_{mse}$ or $\mathcal{L}_{align}$ is present, and the model's performance peaks when both losses are incorporated. 
These findings indicate the indispensability of each component and loss in KnIFE for enhancing the model's domain generalization capability. 
Notebly, DK, Aug and $\mathcal{L}_{mse}$ exert the most significant impact on improvement. These components all contribute to phase-invariant features extraction. 
This emphasizes the critical role of phase-invariant features in EEG generalization.

% Table generated by Excel2LaTeX from sheet 'OpenBMI'
\begin{table}[ht]
  \centering
  \caption{Ablation study on losses (Acc \%).}
    \setlength{\tabcolsep}{3pt}
    \begin{tabular}{cccccc}
    \toprule
    Models & $\mathcal{L}_{mse}$ & $\mathcal{L}_{align}$ & BCICIV-2a & OpenBMI & BCICIV-2b \\
    \midrule
    Baseline & ×     & ×     & 68.02 & 80.16 & 80.48 \\
    Scenario 1 &  \checkmark & ×     & 69.71(+1.69) & 82.82(+2.66) & 80.80(+0.32) \\
    Scenario 2 & ×     &  \checkmark & 69.06(+1.04)& 82.19(+2.03) & 81.28(+0.80)\\
    KnIFE (ours) &  \checkmark &  \checkmark  & \textbf{71.37}(+3.35) & \textbf{83.00}(+2.84) & \textbf{82.53}(+2.05) \\
    \bottomrule
    \end{tabular}%
  \label{loss_ablation}%
\end{table}%

\subsection{Parameter Sensitivity Analysis}
\begin{figure*}[t]
  \centering
  \includegraphics[scale=1]{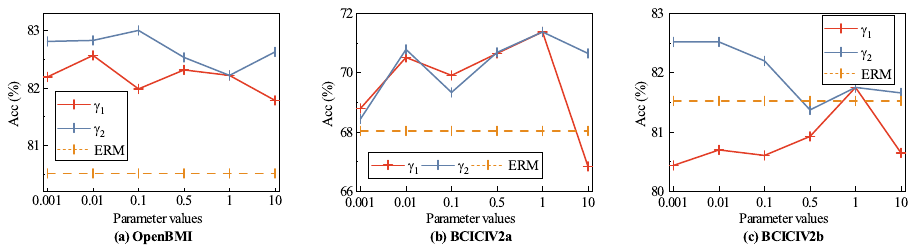}
  \caption{Parameter sensitivity analysis of KnIFE.}
  \label{parameter_analysis}
\end{figure*}

There are two parameters in KnIFE that affect the final loss: $\gamma_1$, which affects the knowledge distillation of phase-invariant feature, and $\gamma_2$, which affects the alignment of inter-domain invariant features. 
Fig. \ref{parameter_analysis} presents the sensitivity analysis of KnIFE's parameters on the three datasets. 
We evaluated it by fixing one parameter and varying the other. 
The results consistently show that KnIFE outperforms the baseline algorithm ERM in almost all cases. An exception is observed when $\gamma_1$ is equal to 10 on the BCICIV2a dataset, where KnIFE shows slightly lower performance than ERM. 
On BCICIV2b dataset, $\gamma_2$ has a better sensitivity on the results than $\gamma_1$. The possible reason is that the dataset has only three channels, which is not conducive to distilling phase-invariant features. 
Overall, these results suggest that KnIFE is, to some extent, insensitive to the choice of these two parameters.

\section{Discussion}
\label{dis}
In this section, we further discuss the positive effect of phase information on our proposed KnIFE, and investigate why phase information and spectrum transfer contribute to domain generalization in the EEG data. overall, we have made four noteworthy observations: 1) Extracting invariant features from both inter- and intra-domain perspectives provides advantages over other diversified perspectives. 2) Compared to amplitude information and raw data, focusing on phase information enhances the model's generalization effect. Additionally, preserving certain amplitude information also proves beneficial. 3) Despite significant disparities in data distribution between subjects, phase information inherently contains more similarities, aiding in the extraction of domain-invariant features. 4) Spectrum transfer preserves the key features of EEG data, explaining why it enhances data similarity without compromising recognition accuracy.

\subsection{Comparison with Recent Methods}
% Table generated by Excel2LaTeX from sheet 'Sheet2'
\begin{table}[t]
  \centering
  \caption{Comparison with state-of-the-art methods in recent years.}
    \begin{tabular}{cccc}
    \toprule
    Datasets & Methods & Year  & Accuracy (\%) \\
    \midrule
    \multirow{3}[2]{*}{BCICIV-2a} & RSC \cite{huang2020self}  & 2020  & 69.29  \\
          & VERx \cite{krueger2021out} & 2021  & 68.40  \\
          & \textbf{KnIFE (ours)} & \textbf{2024} & \textbf{71.37} \\
    \midrule
    \multirow{3}[2]{*}{BCICIV-2b} & MIDRL \cite{jeon2023mutual} & 2023  & 74.46  \\
          & FDCL \cite{liang2023domain} & 2023  & 77.36  \\
          & \textbf{KnIFE (ours)} & \textbf{2024} & \textbf{82.53} \\
    \midrule
    \multirow{4}[2]{*}{OpenBMI} & MIN2NET \cite{9658165} & 2022  & 61.50  \\
          & EEGConformer \cite{9991178} & 2023  & 67.90  \\
          & DRLF \cite{han2024noise} & 2024  & 77.40  \\
          & \textbf{KnIFE (ours)} & \textbf{2024} & \textbf{83.00} \\
    \bottomrule
    \end{tabular}%
  \label{RecentMethods}%
\end{table}%

We compare KnIFE with competitive state-of-the-art methods in recent years, as shown in Table \ref{RecentMethods}.
RSC and Verx, which aim to reduce distribution drift between domains, focus on extracting inter-domain invariant features in EEG decoding. Our KnIFE outperforms them in accuracy by an average of about 2.71\%. 
MIDRL and FDCL both generalize EEG decoding through diverse perspectives. 
MIDRL focuse on class-relevant and subject-invariant feature representations, while FDCL emphasizes category-oriented feature decorrelation and cross-view invariant feature learning. 
KnIFE achieves an average accuracy about 6.62\% higher than theirs.
MIN2NET and EEGConformer are the latest neural networks for EEG decoding. They rely on the network's powerful feature extraction capabilities to achieve a certain generalization effect. However, they lack the ability to learn domain-invariant features, resulting in inferior generalization performance compared to specialized domain generalization methods. 
DRLF disentangle the representation of EEG data into three components to extend the generalization ability of subject-specific features. 
The subject-specific feature is essentially an inter-domain invariant feature, and KnIFE surpasses DRLF by 5.6\% in accuracy.

In summary, solely learning inter-domain invariant features is evidently insufficient for enhancing domain generalization, consistent with the suggestions in references \cite{bui2021exploiting} \cite{zhao2019learning}. KnIFE gains an advantage by learning invariant features from both inter- and intra-domain perspectives compared to other diversified perspectives. The reason may lie in the fact that inter- and intra-domain perspectives are mutually exclusive. Features learned from these two perspectives are expected to have minimal duplication and redundancy, thus containing more valuable information that contributes to domain generalization.

\subsection{Phase information benefits generalization}
\begin{figure}[t]
  \centering
  \includegraphics[scale=1]{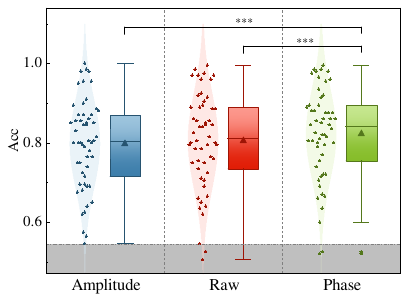}
  \caption{Performances of KnIFE that focuses on different data components. The dots represent the distribution of the subject data, and the triangle is the mean of the data. The gray area contains low-precision subjects.}
  \label{phase_amp}
\end{figure}

The experiments conducted in the preceding sections indicate that emphasizing phase information benefits generalization on unseen domains. To investigate the influence of amplitude and phase information on EEG generalization, we conducted experiments using the large-scale dataset OpenBMI. We modified the KnIFE's teacher network to focus on amplitude information, phase information and raw data respectively. Fig. \ref{phase_amp} presents the outcomes of these three experiments. The phase-focused model achieved the best generalization performance. These results demonstrate that the knowledge distillation framework can extract more robust invariant features from phase information than amplitude information or raw data. Additionally, we made an interesting observation: although phase-focused model exhibited higher accuracy for many subjects compared to the amplitude-focused model, the later outperformed the former for subjects with lower precision. This suggests that amplitude information also contributes to generalization, albeit to a lesser extent. Therefore, it appears prudent to retain a portion of the amplitude information during spectral transfer rather than swapping it entirely.

\subsection{Rationale Behind Effectiveness of Phase Information}
\begin{figure*}
  \centering
  \includegraphics[scale=1]{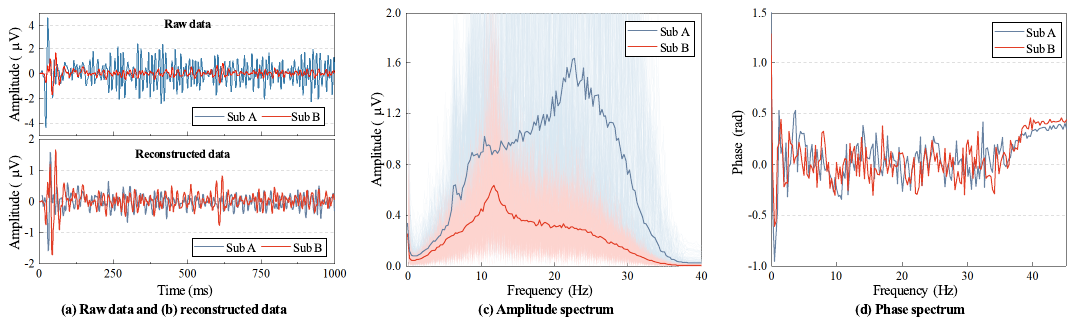}
  \caption{Plots of data from two subjects in the time, amplitude and phase domains. 
  The reconstructed data (b) of subjects A and B exhibit greater similarity than the raw data (a), and their phase spectrum (d) shows more similarity than the amplitude spectrum (c).}
  \label{phase_amp_vis}
\end{figure*}

The effectiveness of phase information in enhancing model generalization performance stems from its capacity to contain more domain-invariant features. 
Here, we delve into the impact of phase information on EEG's domain-invariant feature representation.
Fig. \ref{phase_amp_vis} displays EEG data in the time domain, amplitude spectrum, and phase spectrum from two subjects with markedly different EEG data distributions. 
Table \ref{similarity} quantifies the similarity of the two subjects in Fig. \ref{phase_amp_vis} in terms of Euclidean distance and correlation. 
The larger the Euclidean distance, the lower the similarity; conversely, the greater the correlation coefficient, the higher the similarity.
The data in Fig. \ref{phase_amp_vis}(b) are reconstructed from the raw data in Fig. \ref{phase_amp_vis}(a) using spectral transfer. The similarity between the raw data of subjects A and B is lower than that of the reconstructed data.
This indicates that Fourier-based spectral transfer enhances the similarity of EEG data from disparate subjects. 

Moreover, the amplitude spectrum (c) and phase spectrum (d), derived from the Fourier transformation of the raw data in Fig. \ref{phase_amp_vis}(a), exhibit varying degrees of similarity.
The phase spectrum demonstrates greater similarity than the amplitude spectrum, although their raw data have large distribution differences. 
This observation underscores the robustness of phase information to distribution shifts and its efficacy in representing phase-invariant feature.

In fact, during motor imagery tasks, differences in activation levels across frequency bands such as $\alpha$ (8-13 Hz) and $\beta$ (15-30 Hz) rhythms are apparent. 
For instance, Fig. \ref{phase_amp_vis}(c) illustrates that subject A evokes significantly higher $\beta$ rhythm amplitude compared to subject B. 
Conversely, such differences are less pronounced in the phase domain (Fig. \ref{phase_amp_vis}(d)).
Therefore, it is not difficult to understand that phase information can maintain greater domain invariance than amplitude.

% Table generated by Excel2LaTeX from sheet 'OpenBMI'
\begin{table}[t]
  \centering
  \caption{Quantification of similarity in Fig. \ref{phase_amp_vis} in terms of distance and correlation.}
    \begin{tabular}{ccccc}
    \toprule
    Similarity metrics & (a)   & (b)   & (c)   & (d) \\
    \midrule
    Euclidean  & 29.92 & 11.35 & 4045  & 2.813 \\
    Correlation   & 0.8914 & 0.892 & 0.2101 & 0.3115 \\
    \bottomrule
    \end{tabular}%
  \label{similarity}%
\end{table}%

\subsection{Discussion of Spectral Transfer}
\begin{figure}
  \centering
  \includegraphics[scale=1]{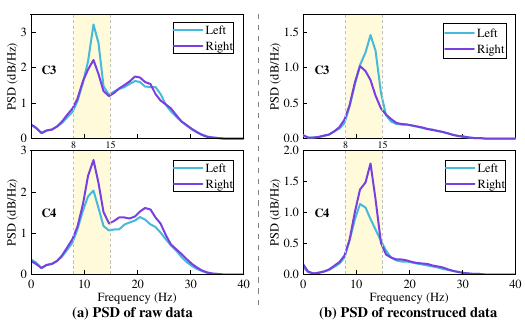}
  \caption{Visualization of ERD/ERS patterns by PSD for MI data. The yellow area is 8-15 Hz frequency band. The upper row displays figures for channel C3, while the bottom row shows figures for channel C4.}
  \label{PSD}
\end{figure}

Spectral transfer focus on phase information by preserving the phase spectrum of different domains and swapping the amplitude spectrum. Does this transformation compromise the fundamental characteristics of EEG data? We address this question by analyzing the event-related desynchronization (ERD) and event-related synchronization (ERS) phenomenon in motor imagery tasks. Pfurtscheller et al. \cite{pfurtscheller1996event} found that during performing motor imagery, there is a decrease in the amplitude of the $\alpha$ frequency band (8-13 Hz) in the contralateral sensorimotor cortex's EEG signal, known as ERD, indicating a reduction in the amplitude of the activated cortical EEG signal. Simultaneously,  there is an increase in the amplitude of the $\alpha$ frequency band in the ipsilateral sensorimotor cortex signal, termed ERS, signifying an increase in the amplitude of the corresponding cortex during the resting state. The ERD/ERS phenomenon is a crucial feature in motor imagery recognition. Fig. \ref{PSD} illustrates the power spectral density (PSD) on channels C3 and C4 for both raw and reconstructed data. It can be observed that both the raw and reconstructed data exhibit the ERD/ERS phenomenon. This observation suggests that spectral transfer does not disrupt the fundamental characteristics of EEG. It reinforces the notion that spectral transfer enhances data similarity by emphasizing phase features without compromising the model's classification performance.

\subsection{limitations and Future Directions}
The spectral transfer and knowledge distillation methods employed by KnIFE rely on Fourier transform techniques, necessitating direct input of raw EEG data rather than preprossed  feature vectors. 
Consequently, KnIFE's efficacy is closely tied to the choice of deep neural network used as its backbone. 
In this study, we employed classic neural networks like EEGNet as the backbone. 
Recent advancements have shown promising outcomes in integrating attention mechanisms into EEG decoding networks. In future work, we plan to explore the suitability of these novel neural networks as backbones for EEG domain generalization.
Moreover, there are several aspects of spectral transfer in EEG data that merit further investigation. Specifically, when swapping amplitude spectra, it is crucial to explore how to judiciously select the frequency range. Additionally, constructing an intermediate amplitude spectrum by interpolating the amplitude spectra of two domains presents an feasible way for future research.

\section{Conclusion}
In this paper, we propose a zero-calibration domain-generalization network for EEG decoding, named KnIFE. 
KnIFE is trained using pre-collected data and can be applied directly to new subjects without calibration. 
Its strong generalization capability stems from learning both intra-domain and inter-domain invariant features. 
Phase information serves as the intra-domain invariant features, while correlation alignment is employed to extract inter-domain invariant features. 
We enhance the phase information through spectral transfer and extract phase-invariant features using knowledge distillation. 
Extensive experiments on multiple datasets demonstrate KnIFE's state-of-the-art performance.
Furthermore, we delve into the possible reasons why phase information contributes to EEG domain-invariant feature extraction.
We hope that our approach will inspire further research in EEG domain generalization, particularly 
methods that exploit phase information.

\section*{References}
\balance
\bibliographystyle{ieeetr}
\bibliography{references}
\end{document}